\documentclass[prd,a4paper,preprint,preprintnumbers,nofootinbib]{revtex4-2}

\usepackage{amsmath,amssymb}
\usepackage{graphicx,multirow}
\usepackage{subcaption}
\usepackage[font=small,labelfont=bf]{caption}
\usepackage{tabularx}
\usepackage{slashed}
\usepackage{units}
\usepackage[hyperfootnotes=false]{hyperref}
\usepackage[usenames, dvipsnames]{color}

\newcommand{\bea}{\begin{eqnarray}}
\newcommand{\eea}{\end{eqnarray}}
\newcommand{\beq}{\begin{eqnarray}}
\newcommand{\eeq}{\end{eqnarray}}

\allowdisplaybreaks

\begin{document}

\preprint{UCI-TR-2023-10}

\title{On Q-balls in Anti de Sitter Space}

\author{Arvind~Rajaraman}
\email{arajaram@uci.edu}
\affiliation{Department of Physics and Astronomy, 
University of California, Irvine, CA 92697-4575, USA
}

\author{Alexander~Stewart}
\email{ajstewa1@uci.edu}
\affiliation{Department of Physics and Astronomy, 
University of California, Irvine, CA 92697-4575, USA
}

\author{Christopher~B.~Verhaaren}
\email{verhaaren@physics.byu.edu}
\affiliation{Department of Physics and Astronomy, Brigham Young University, Provo, UT, 84602, USA
}

\date{\today}

\begin{abstract}
We perform a general analysis of thin-wall Q-balls in AdS space. We provide numeric solutions and highly accurate analytic approximations over much of the parameter space. These analytic solutions show that AdS Q-balls exhibit significant differences from the corresponding flat space solitons. This includes having a maximum radius beyond which the Q-balls are unstable to a new type of state where the Q-ball coexists with a gas of massive particles. The phase transition to this novel state is found to be a zero-temperature third-order transition. This, through the AdS/CFT correspondence, has implications for a scalar condensate in the boundary theory. 
\end{abstract}


\maketitle

\tableofcontents

\section{Introduction}
The AdS/CFT correspondence~\cite{Aharony:1999ti}, which relates a theory in an anti-de-Sitter (AdS) space  in $D+1$ dimensions with a conformal field theory in $D$ dimensions,  has been the subject of intense study for many years now.  Any observable quantity in the gravitational theory (sometimes called the bulk theory) has a corresponding quantity in the boundary theory e.g. the 
correlation functions of the field theory are in one-to-one correspondence with the boundary-to-boundary propagators in the bulk theory.
The original correspondence of~\cite{Maldacena:1997re} has since been generalized to many other theories.

When the field theory is at nonzero temperature, it can exhibit phase transitions. In the bulk theory, these often correspond to a formation or alteration of bound states. For example, when the temperature of the field theory is increased, there is a deconfinement phase transition; this is believed to correspond to the formation of a black hole~\cite{Witten:1998zw}. 
AdS can also support other solitons such as boson stars, compact objects made of scalar fields with  gravitational (and potentially gauge) interactions
~\cite{Astefanesei:2003qy,Prikas:2004yw,Nogueira:2013if,Brihaye:2013hx,Buchel:2013uba,Kichakova:2013sza,Kumar:2016sxx,Brihaye:2022oaf}.
 These solitons exhibit yet other types of phenomena; for instance, if one considers charged boson stars, one can find a zero-temperature second order phase transition between two types of boson stars
\cite{Hu:2012dx}.

In this work we consider a different type of soliton in AdS. 
Q-balls are configurations of complex scalars $\phi$ that are bound together by self-interactions~\cite{Coleman:1985ki} (Q-balls can also carry a $U(1)$ gauge charge, but here we consider Q-balls whose binding energy is dominated by the scalar attraction). They are stable over a large range of parameter space, and are hence an example of a nontopological soliton.

As Q-balls are solitons in a scalar field theory, their profiles are the solutions to a nonlinear differential equation. While this differential equation is difficult to solve exactly, extremely accurate approximation methods have been found for Q-balls with thin-wall configurations in~\cite{Heeck:2020bau}. These methods produce excellent analytic approximations to the charge and energy of both global~\cite{Heeck:2020bau} and gauged Q-balls \cite{Heeck:2021zvk} and Q-shells~\cite{Heeck:2021bce, Heeck:2021gam}, as well as the radial excitations of global Q-balls~\cite{Almumin:2021gax}.

Here we extend these methods to study Q-balls in AdS (some previous work on this subject is~\cite{Hartmann:2012wa,Hartmann:2013kna,Kichakova:2013sza}).
 We ignore backreaction on the spacetime and hence the AdS space is taken to be a fixed background for the scalar field. We 
find solutions both numerically and by an extension of the methods of~\cite{Heeck:2020bau}, and show excellent agreement between the approximate analytical functions and the exact numerical solutions in several different spatial dimensions. These methods
agree in showing that there is generically a maximal radius for Q-balls in AdS, except for a special class of scalar potentials such as those studied in~\cite{Hartmann:2012wa}.
Beyond the corresponding maximal charge, we find an unexpected phase of the theory, where the Q-ball coexists with a gas of noninteracting massive particles (a similar phenomenon has been found for Kerr black holes in AdS~\cite{Kim:2023sig}). 

We also examine the consequences of the Q-ball physics for the dual theory on the boundary. We find an intricate phase diagram with zero-temperature phase transitions that can be either second-order, or surprisingly, even third order. These high order phase transitions are found to be generic for a large class of potentials, as long as they admit Q-ball solutions. These results are seen to hold for dual theories in several dimensions.
 
 In the next section we review 
the action for Q-balls in AdS and find the equations satisfied by a spherical Q-balls. We show that thermodynamic relation $dE=\omega dQ$ between the energy and charge holds for AdS solitons. 
We then apply approximation techniques to find a relation between the radius of the Q-ball and the parameters of the theory in Sec.~\ref{s.thinwall}. In Sec.~\ref{s.Numerics} we compare with numerical results, showing remarkable agreement for thin-wall solitons. The numerical results also confirm the analytic prediction that, unlike in flat space, there are soliton solutions with $\omega>m_\phi$. The consequences of our analytic understanding of Q-balls are further explored in Secs.~\ref{s.instab} and~\ref{s.LR} with particular interest given to soliton instabilities and that, in contrast to flat space, the AdS Q-balls often have a maximum allowed radius. The implications of these results for the holographic theory on the boundary are discussed in Sec.~\ref{s.CFT}. In Sec.~\ref{s.nDimAdsQ} we show that the results obtained for 3-dimensional AdS are easily extended to higher and lower dimensions. We close with a discussion of our results.

\section{Q-balls in Anti-de Sitter\label{s.AdsQ}}
We consider a complex scalar field propagating in a fixed Anti-de Sitter background.
The AdS geometry is parameterized as
\beq
ds^2=a(r)dt^2-b(r)dr^2-r^2\left(d\theta^2+\sin^2\theta d\phi^2\right)~,
\eeq
with
\beq
a(r)=1+\frac{r^2}{\ell^2}~, \ \ b(r)=\frac{1}{1+\frac{r^2}{\ell^2}}~,\label{e.AdSgrav}
\eeq
where $\ell = \sqrt{-\frac{3}{\Lambda}}$ is a scale that sets the size of the AdS space. 

 The action for a scalar field propagating in this geometry is
\beq
\mathcal{S}=\int d^4x\sqrt{-g}\left[\left(\nabla_\mu\Phi \right)^\ast\left(\nabla^\mu\Phi \right)-U(\Phi\Phi^\ast) \right]~,
\eeq
The complex scalar field $\Phi$ is subject to a potential $U(\Phi\Phi^\ast) $ that preserves a global $U(1)$ symmetry and thereby leads to a conserved particle number, denoted by $Q$
\beq
Q=i\int d^3x\sqrt{-g}\left(\Phi^\ast\partial^t\Phi-\Phi\partial^t\Phi^\ast \right)~.
\eeq
The energy stored in the field
\beq
E=\int d^3x\sqrt{-g}\,T^t_{\phantom{t}t}~,
\eeq
 is given in terms of the energy-momentum tensor
 \beq
T_{\mu\nu}=\left(\nabla_\mu\Phi \right)^\ast\left(\nabla_\nu\Phi \right)+\left(\nabla_\nu\Phi \right)^\ast\left(\nabla_\mu\Phi \right)-g_{\mu\nu}\left[\left(\nabla_\mu\Phi \right)^\ast\left(\nabla^\mu\Phi \right)-U(\Phi\Phi^\ast) \right]~.
 \eeq

The equation for $\Phi$ is
\beq
\nabla_\mu\nabla^\mu\Phi=-\frac{\partial U}{\partial \Phi^\ast}=-\Phi \frac{d U}{d(\Phi\Phi^\ast)}~.
\eeq
We look for spherical Q-ball solutions; this implies that we take  the scalar field to be of the form
\beq
\Phi=\frac{\phi_0}{\sqrt{2}}f(r)e^{-i\omega t}~,
\eeq
where $\phi_0/\sqrt{2}$ is the value of the nontrivial minimum of $U(\Phi\Phi^\ast)/(\Phi\Phi^\ast)$. To make connection with the typical flat space characterization of Q-balls we also define
\beq
\omega_0^2\equiv\left. \frac{U(\Phi\Phi^\ast)}{\Phi\Phi^\ast}\right|_{\Phi=\phi_0/\sqrt{2}}~.
\eeq
 The dimensionless function $f(r)$ is referred to as the profile of the field configuration. 
The equation of motion for the field can then be written as an equation to determine this profile
\beq
{d^2f\over dr^2}+\frac{2}{r}\frac{1+2r^2/\ell^2}{1+r^2/\ell^2}{df\over dr}=\frac{f}{1+r^2/\ell^2}\left(\frac{d U}{d(\Phi\Phi^\ast)}-\frac{\omega^2}{1+r^2/\ell^2} \right)~.
\eeq
Note that as $\ell\to\infty$ this becomes the usual flat space equation for $f$~\cite{Coleman:1985ki}.

In general, the equation for $f$ cannot be solved analytically. In this study we focus, following~\cite{Heeck:2020bau}, on a sextic potential
\beq
U(|\phi|)=m_\phi^2|\phi|^2-\beta|\phi|^4+\frac{\xi}{m_\phi^2}|\phi|^6~.
\eeq
While a specific choice, this potential captures the qualities of many potentials that give rise to Q-balls~\cite{Heeck:2022iky}. In particular, we expect that any potential that gives rise to thin-wall Q-balls will have qualitatively similar results. In this potential we have used the mass of the scalar field to determine the estimated scaling of higher dimension operators. Our assumption of the scalar field not back reacting on the spacetime is essentially that $m_\phi$ and $\phi_0$ are much smaller than the Planck scale. We also assume that cosmological constant $\Lambda$ is much smaller than $m_\phi$ to ensure that $\ell^2|\phi|^6$ contributions to the potential are subleading. This leads to the constraint $\ell m_\phi\gg1$. 

The potential parameters given above are mapped to the Q-ball parameters $\phi_0$ and $\omega_0$ by
\beq
\phi_0=m_\phi\sqrt{\frac{\beta}{\xi}}~, \ \ \ \ \omega_0=m_\phi\sqrt{1-\frac{\beta^2}{4\xi}}~.
\eeq
It is convenient for both numerical analyses and in determining how various quantities depend on the parameters of the theory to define the dimensionless quantities
\bea
\Phi_0=\frac{\phi_0}{\sqrt{m_\phi^2-\omega_0^2}}~, \ \ \ \ \Omega_{(0)}=\frac{\omega_{(0)}}{\sqrt{m_\phi^2-\omega_0^2}}~, \ \ \ \ 
\rho=r\sqrt{m_\phi^2-\omega_0^2}~, \ \ \ \ \lambda=\ell\sqrt{m_\phi^2-\omega_0^2}~.
\eea
We see that our $\ell m_\phi\gg1$ constraint becomes $\lambda\gg1$. 
The $f$ equation can then be written as
\begin{align}
f''+\frac{2}{\rho}\frac{1+2\rho^2/\lambda^2}{1+\rho^2/\lambda^2}f'
&=\frac{f}{1+\rho^2/\lambda^2}\left[1-\frac{\Omega^2}{1+\rho^2/\lambda^2}+\Omega_0^2-4f^2+3f^4\right]~,
\label{eq:eom}
\end{align}
while the charge is
\bea
Q = 4\pi \Phi_{0}^{2}\Omega
\int_{0}^{\infty}\mathrm{d}\rho \rho^{2} \frac{f^{2}}{1+\rho^{2}/\lambda^{2}} ~,
\eea
and the energy is
\begin{align}
E
=& \sqrt{m_\phi^2-\omega_0^2}2\pi\Phi_{0}^{2}\int_{0}^{\infty}\mathrm{d}\rho \rho^{2}\left(\frac{\Omega^{2}f^{2}}{1+\rho^{2}/\lambda^{2}}+f^{\prime 2}(1+\rho^{2}\lambda^{2})+f^{2}(1-f^{2})^{2}+\Omega_{0}^{2}f^{2}\right)~\nonumber\\
=&\omega Q-L~,
\end{align}
where the Lagrangian is given by
\beq
L=\sqrt{m_\phi^2-\omega_0^2}2\pi\Phi_{0}^{2}\int_{0}^{\infty}\mathrm{d}\rho \rho^{2}\left(\frac{\Omega^{2}f^{2}}{1+\rho^{2}/\lambda^{2}}-f^{\prime 2}(1+\rho^{2}\lambda^{2})-f^{2}(1-f^{2})^{2}-\Omega_{0}^{2}f^{2}\right)~.
\eeq

Using the Lagrangian, one finds that field configurations that satisfy the equations of motion also satisfy
\beq
\frac{dL}{d\omega}=Q~,
\eeq
as in flat space. A straightforward calculation produces
\beq
\frac{dE}{d\omega}=Q+\omega\frac{dQ}{d\omega}-\frac{dL}{d\omega}=\omega\frac{dQ}{d\omega}~,
\eeq
implying that
\beq
\frac{dE}{dQ}=\omega~.
\eeq
We emphasize that although we have chosen a specific potential, this differential relation relating the energy and charge holds for all potentials. This result also gives $\omega$ a simple interpretation as a chemical potential, describing how the energy changes as the particle number changes.

\section{Thin-Wall Q-balls\label{s.thinwall}}

Following Coleman~\cite{Coleman:1985ki}, we can treat Eq.~\eqref{eq:eom} above as describing a particle rolling in a potential $V(f,\rho)$ 
defined as 
\begin{align}
V(f,\rho)=\frac{f^2}{2\left(1+\rho^2/\lambda^2 \right)}\left[\frac{\Omega^2}{1+\rho^2/\lambda^2}-\Omega_0^2-(1-f^2)^2 \right]~,
\end{align}
where the $\rho$ coordinate is treated as an effective time. 
Since the potential depends on $\rho$, the physics is of a particle rolling in a time dependent potential. In flat space and in AdS the coefficient of the $f'$ term decreases with increasing $\rho$, which is interpreted as a friction that decreases with time. In flat space, thin-wall trajectories are those in which the particle starts rolling from rest near a maximum in the potential but does not complete the large field change, rolling down the hill, until the friction term has become somewhat small. This leads to a fast transition, or a thin-wall for the soliton. All of the Q-ball trajectories end at $f=0$ to ensure a field configuration that is localized in space. 

\begin{figure}
\centering
\begin{subfigure}{.5\textwidth}
  \centering
  \includegraphics[width=.8\linewidth]{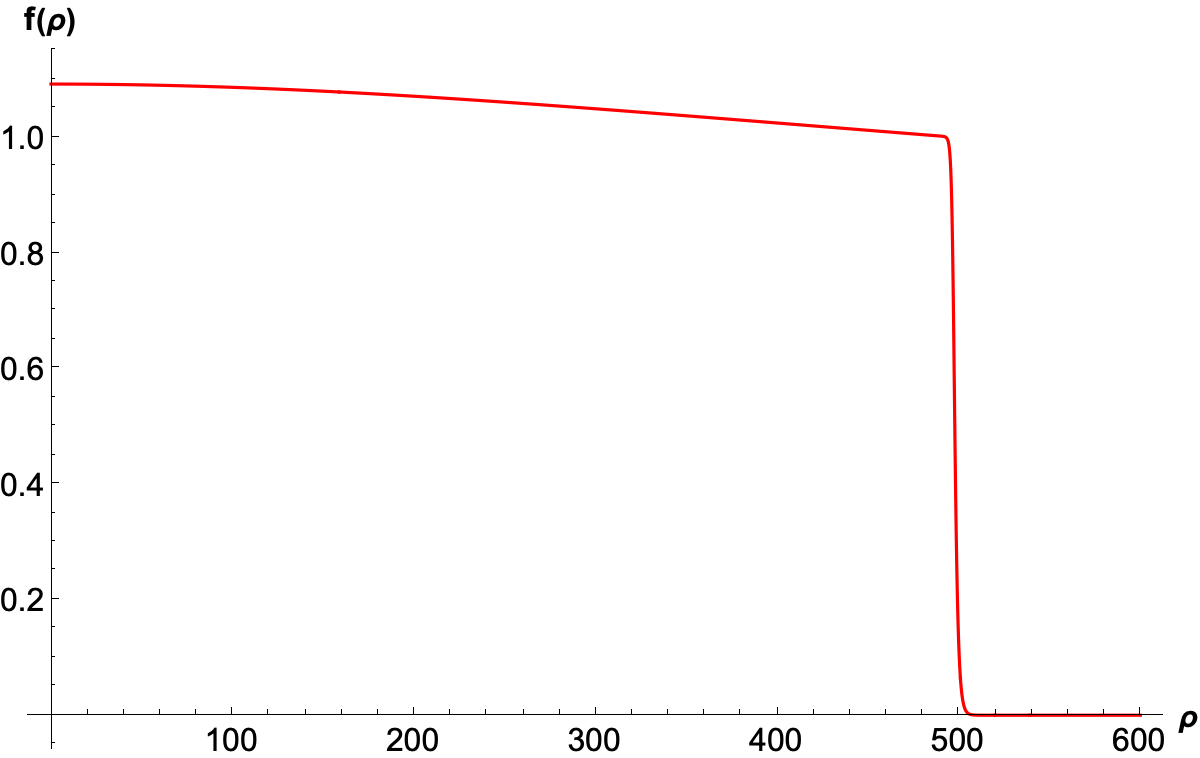}
  \label{fig:sub1_3}
\end{subfigure}%
\begin{subfigure}{.5\textwidth}
  \centering
  \includegraphics[width=.8\linewidth]{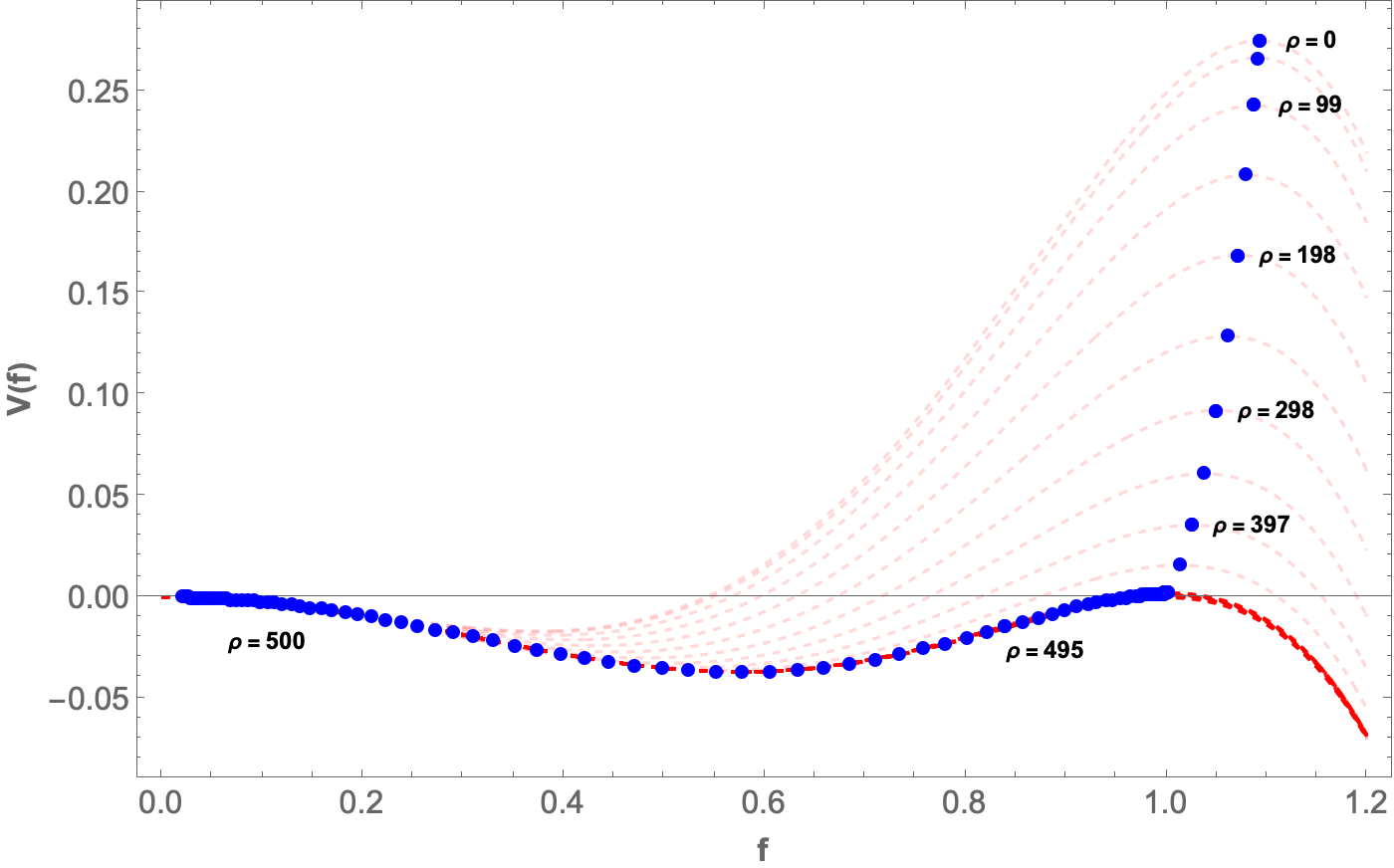}
  \label{fig:sub2_3}
\end{subfigure}
\caption{
On left: Numerical Q-ball profile for $\lambda = 500$ and $\Omega_{0} = 1/\sqrt{2}$ , $\Omega=0.99$. On right: effective potential $V(f,\rho)$ for the same parameters and various values of $\rho$, where the blue dots indicate the location of the field value.}
\label{fig:exampleprof}
\end{figure}

An example of thin-wall Q-balls in AdS, where the field stays at the false minimum of $U$ at $f=f_+$ for a long time before making a quick transition to the true minimum of $U$ at $f=0$, is shown in Fig.~\ref{fig:exampleprof}. The left panel shows the Q-ball profile found numerically using  collocation algorithms implemented by SciPy~\cite{2020SciPy-NMeth}.
On the right panel, we have shown the effective potential $V(f,\rho)$ as a function of $f$ for several values of $\rho$. The field value at specific value of $\rho$ is shown as a dot. 

The effective potential is found to have extrema at $f=0$ and at
\bea
f^2_\pm\equiv\frac23\pm \frac13\sqrt{1+\frac{3\Omega^2}{1+\rho^2/\lambda^2}-3\Omega_0^2}~.
\eea
We note that $f_+$ can take real values while $f_-$ does not when
\beq
\Omega^2\geq\left(1+\frac{\rho^2}{\lambda^2} \right)\left(1+\Omega_0^2 \right)~.
\eeq
Within those restrictions we find that $f_+$ corresponds to a maximum and $f_-$ to a minimum.  A more restrictive constraint on Q-ball profiles is that if the ``particle" is to roll down a slope that ends at the $f_+$ maximum then that maximum must have positive energy, or $V(f,\rho)>0$, for the particle trajectory to overcome the residual friction and end at $f=0$. However, we find that $V(f_+(\rho),\rho)=0$ for
\beq
\rho^2=\lambda^2\frac{\Omega^2-\Omega_0^2}{\Omega_0^2}~.
\eeq
Trajectories that have not already rolled to near the $f=0$ maximum by this value of $\rho$ cannot produce a localized soliton configuration. This limits the radius of thin-wall AdS Q-balls to
\beq
R_\text{thin-wall}\lesssim\lambda\frac{\sqrt{\Omega^2-\Omega_0^2}}{\Omega_0}~.\label{e.Rads}
\eeq

For flat-space Q-balls the extremum at $f=0$ is always a maximum because $\Omega^2-\Omega_0^2<1$ for Q-ball configurations. In an AdS background, as we see below, this restriction does not apply. Thus, we find that
\beq
\left.\frac{d^2V}{df^2}\right|_{f=0}=\frac{\displaystyle\Omega^2-\left(1+\rho^2/\lambda^2 \right)\left(1+\Omega_0^2 \right)}{\left(1+\rho^2/\lambda^2 \right)^2}~.
\eeq
can lead to a minimum at $f=0$, at least until $\rho$ becomes sufficiently large. Note that this is exactly the same condition for the minimum at $f_-$ to not exist. In this case the maximum at $f_+$ rolls directly to the minimum at $f=0$. This behavior gives rise to a class of AdS Q-balls with no flat space analogue.

We take the large field transition to occurs at  $\rho=R\gg 1$. This implies that
for $\rho\ll R$, we should take the solution to be approximately $f=f_+(\rho)$. This
is not an exact solution to the equations, because the derivatives of $f_+(\rho)$ are nonzero, but
 $f'_+/f_+$ is of order ${1\over \lambda}$, so for $\lambda\gg1$ the corrections to this solution are small.

On the other hand, when $\rho\sim R$ the field transitions between the two vacua, and the
derivatives are not small. 
To analyze this region, we define an energy-like quantity
\begin{align}
W=&\frac12f^{\prime2}+
V(f,\rho)~,
\label{eq:edef}
\end{align}
to analyze the transition region of the Q-ball. For a thin-wall transition we expect this ``energy" to be approximately conserved, so $W$ is constant. Note that as $\rho\to\infty$ that $W\to0$ because the particle comes to rest at $V(0,\infty)=0$. In this approximation we can neglect the contributions to $f'$ away from the radius of the soliton, which we define by $f''(R)=0$. The equation is then
\beq
\frac{df}{d\rho}=\sqrt{-2V(f,R)}=\frac{f}{\sqrt{1+R^2/\lambda^2}}\sqrt{(1-f^2)^2-\frac{\Omega^2}{1+R^2/\lambda^2}+\Omega_0^2 }~.
\eeq
Finally, we expect this equation to be most correct when the particle rolls with the least amount of friction. This would be for cases that come as close as possible to saturating the bound in Eq.~\eqref{e.Rads}. In this case the differential equation is
\beq
\frac{df}{d\rho}=\frac{f}{\sqrt{1+R^2/\lambda^2}}(1-f^2)~,
\eeq
which leads to the transition profile
\beq
f_t(\rho)=\frac{1}{\sqrt{1+2e^{2(\rho-R)/\sqrt{1+R^2/\lambda^2}}}}~.
\eeq
To match the interior solution we multiply by $f_+(\rho)$. So, our approximate thin-wall profile is
\beq
f_T(\rho)=\frac{f_+(\rho)}{\sqrt{1+2e^{2(\rho-R)/\sqrt{1+R^2/\lambda^2}}}}~.\label{eq:transfn}
\eeq

This transition function allows us to capture the leading order effects of friction of the particle trajectories. Using the equations of motion for $f$ we find
\bea
{dW\over d\rho}=-
\frac{2}{\rho}\frac{1+2\rho^2/\lambda^2}{1+\rho^2/\lambda^2}f^{\prime2}
+\frac{\rho f^2}{\lambda^2\left(1+\rho^2/\lambda^2 \right)^2}\left[\Omega_0^2-\frac{2\Omega^2}{1+\rho^2/\lambda^2 }+\left(1-f^2\right)^2 \right]
\eea
By integrating from $\rho=R-z_0$ to $\rho=\infty$ we find
\begin{align}
W_{\infty}-W_{R-z_0}=&-
2\int_{R-z_0}^{\infty}\frac{d\rho}{\rho}\frac{1+2\rho^2/\lambda^2}{1+\rho^2/\lambda^2}f^{\prime2}
\nonumber
\\
&+\int_{R-z_0}^{\infty}d\rho\frac{\rho f^2}{\lambda^2\left(1+\rho^2/\lambda^2 \right)^2}\left[\Omega_0^2-\frac{2\Omega^2}{1+\rho^2/\lambda^2 }+\left(1-f^2\right)^2 \right]
\label{eq:integralreln}
\end{align}
We find the leading effect from the friction by evaluating these integrals using the transition function defined in Eq.~\eqref{eq:transfn}, similar to the analysis done in~\cite{Heeck:2020bau}.
Equation (\ref{eq:integralreln}) then yields the relation
\bea
\Omega_0^2-\frac{\Omega^2}{1+\frac{R^{2}}{\lambda^{2}}}
=\frac{\lambda^{2}R^{2}(\Omega^{2}\ln{4}-5)-2\lambda^{4}-3R^{4}}{2R\lambda(\lambda^{2}+R^{2})^{3/2}}~.
\label{eq:RvWreln}
\eea
This relation for a more general $n$-dimensional AdS background is derived in more detail in Appendix \ref{app:Rderiv}.

\section{Numerics\label{s.Numerics}}
We begin by verifying that our analytic understanding of Q-balls in AdS agrees
with numerical solutions.
Solutions to the profile equation~\eqref{eq:eom} were obtained numerically through the use of  SciPy~\cite{2020SciPy-NMeth}.
We consider two benchmark points, corresponding to (a) $\lambda=500,\, \Omega_0=1/\sqrt{2}$
and (b) $\lambda=100,\, \Omega_0=1/10$. 
For each benchmark point, solutions were generated for various choices of $\Omega$.

\begin{figure}
\centering
\begin{subfigure}{.5\textwidth}
  \centering
  \includegraphics[width=.8\linewidth]{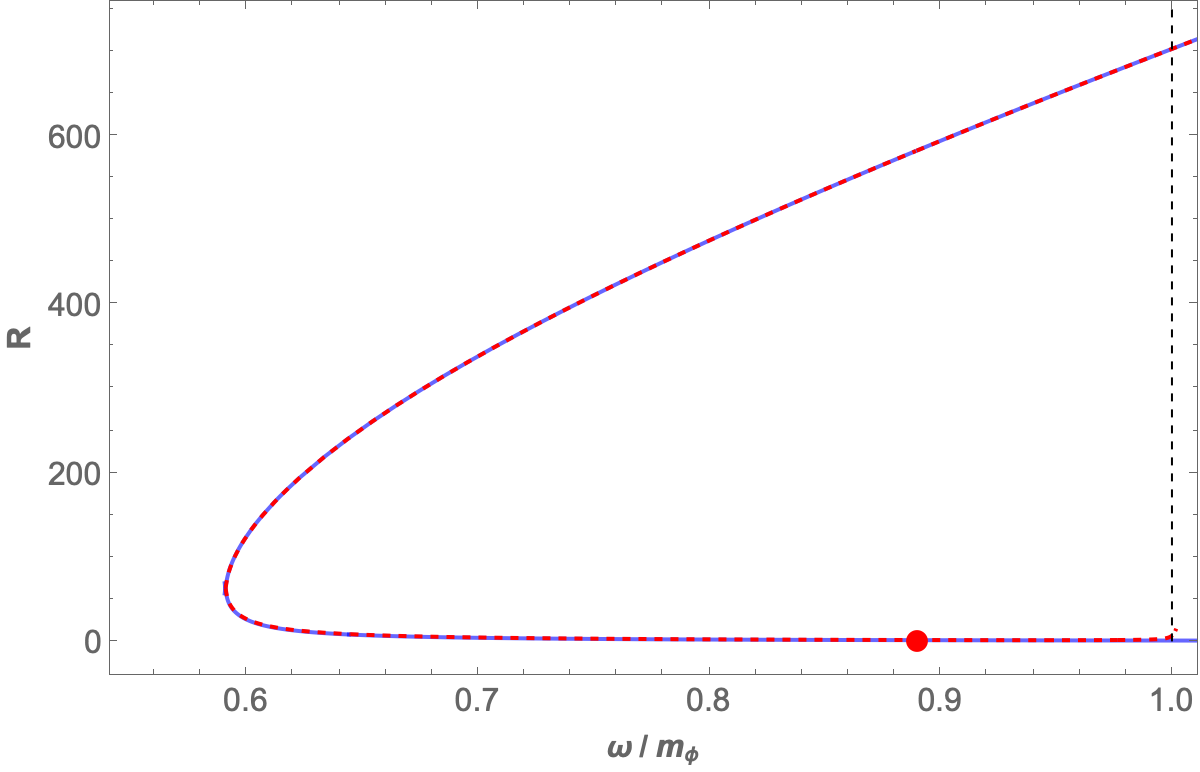}
  \label{fig:sub1_1}
\end{subfigure}%
\begin{subfigure}{.5\textwidth}
  \centering
  \includegraphics[width=.8\linewidth]{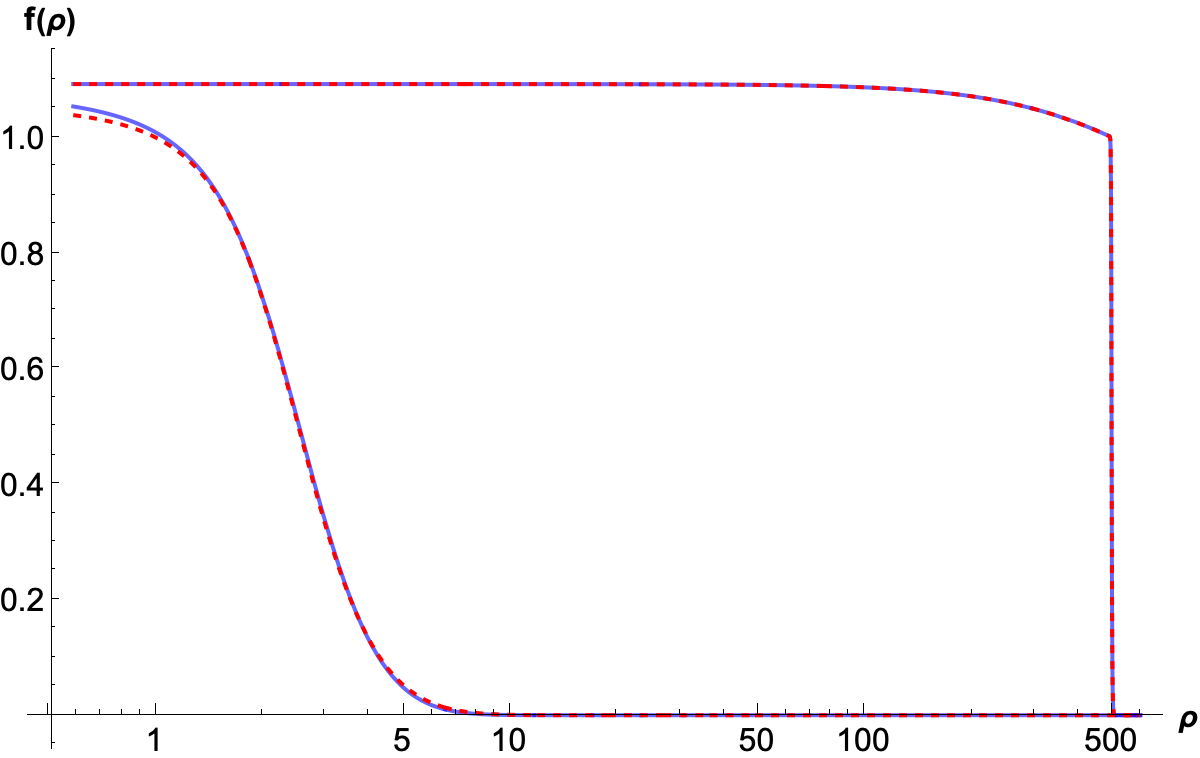}
  \label{fig:sub2_1}
\end{subfigure}%
\caption{Numerical calculations (red dashed) compared to analytic predictions (blue) for $\lambda=500,\, \Omega_0=1/\sqrt{2}$. Left panel: Q-ball radius $R$ as a function of $\omega$. Right panel: Both Q-ball profiles for $\Omega=0.99$.
}
\label{fig:RvW500}
\end{figure}

Figure~\ref{fig:RvW500} shows the comparison between the numerical calculations (red dashed) and analytic predictions (blue) for the benchmark point $\lambda=500, \Omega_0=1/\sqrt{2}$. The left panel shows the obtained numeric values for the Q-ball radius $R$ as a function of $\omega$ with a comparison to the analytic predictions from Eq.~\eqref{eq:RvWreln}.  The difference between the numerical and analytical results is negligible.
The vertical dashed line is $\omega=m_\phi$ or $\Omega=\sqrt{1+\Omega_0^2}$. In flat space there are no Q-ball solutions beyond this point, but we find both numerical and theoretical evidence for these solitons in AdS space. The red dot indicates the $E=mQ$ instability point which is described below. Soliton solutions to the right of this point on the curve are unstable to dissociation.

 The right panel of  Fig.~\ref{fig:RvW500} compares the numerical and analytic predictions (as predicted by (\ref{eq:transfn})) for the two Q-ball profiles for the same parameters $\lambda=500,\, \Omega_0=1/\sqrt{2}$ and $\Omega=0.99$. The two different solutions for $R$ lead to two different profile functions of very different radius.  Once again, the difference between the numerical and analytical results is negligible for both profiles.

 \begin{figure}
\centering
\begin{subfigure}{.5\textwidth}
  \centering
  \includegraphics[width=.8\linewidth]{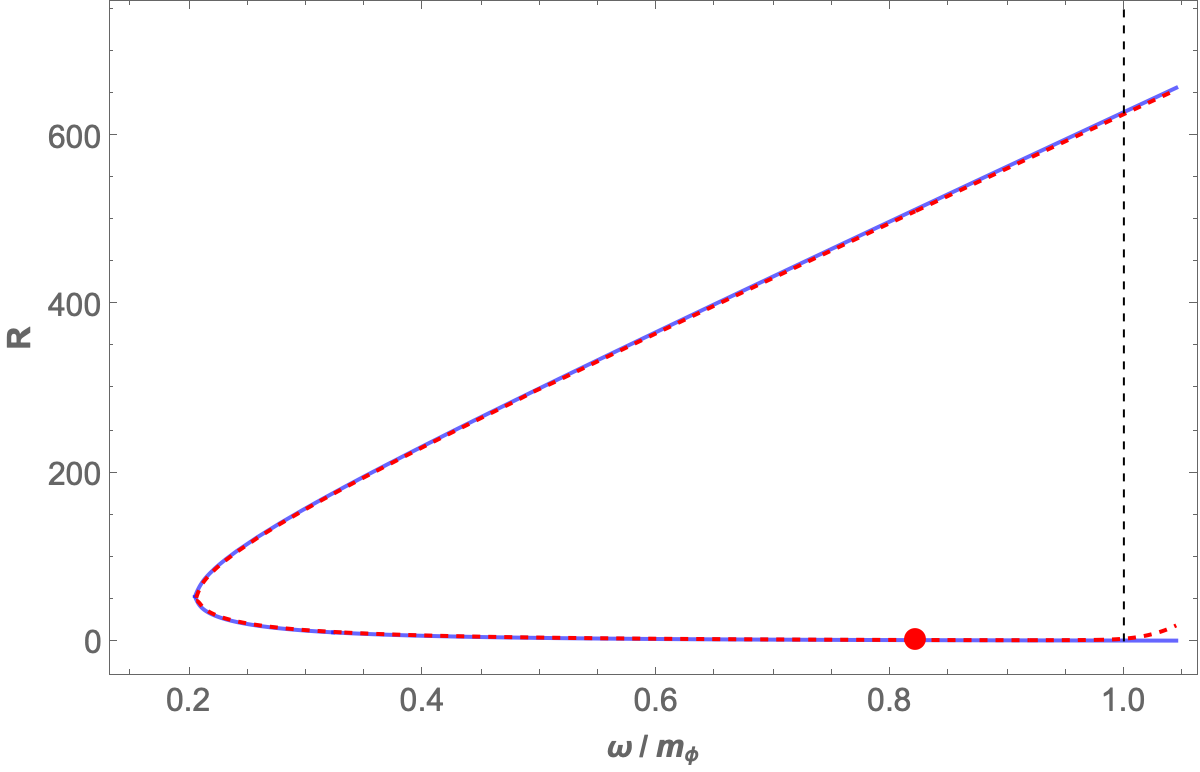}
  \label{fig:sub2_2}
\end{subfigure}
\begin{subfigure}{.49\textwidth}
  \centering
  \includegraphics[width=.8\linewidth]{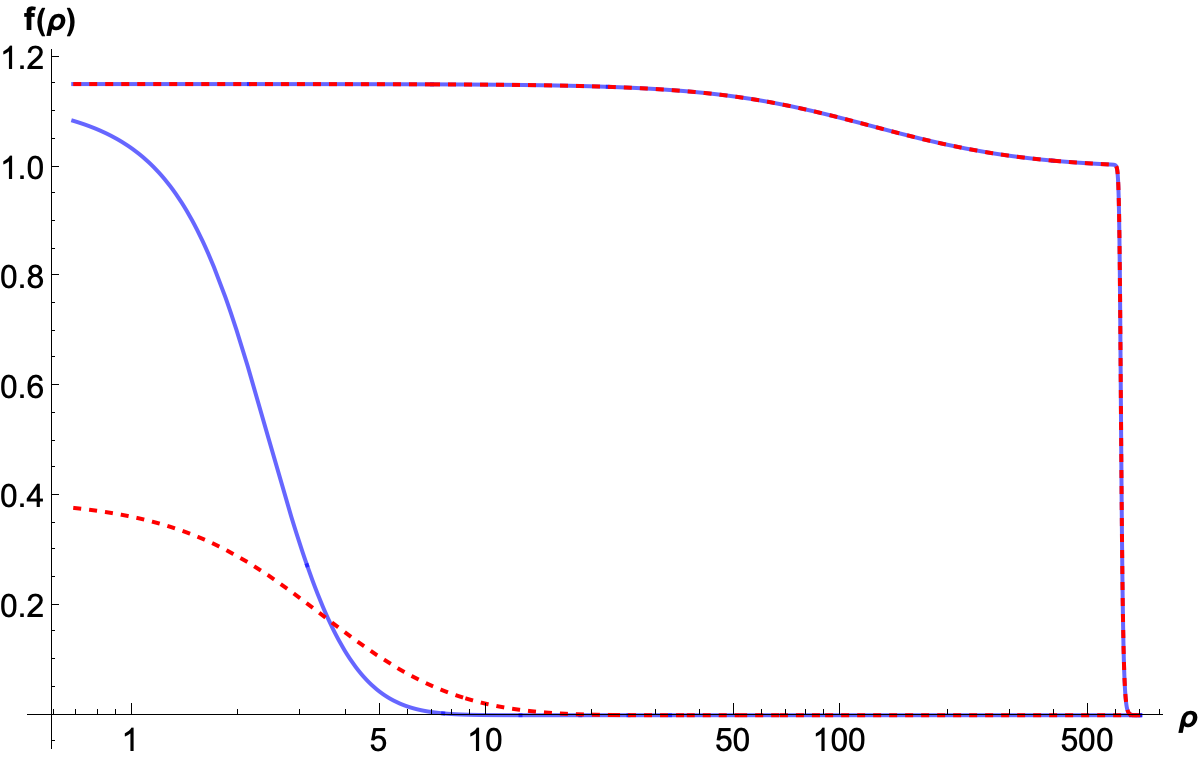}
  \label{fig:sub2_2}
\end{subfigure}%
\caption{As in Fig.~\ref{fig:RvW500}, but for the benchmark point $\lambda=100,\, \Omega_0=1/10$.  Red dashed curves are numerical results, blue curves are the analytic predictions.}
\label{fig:RvW100}
\end{figure}
 
 Figure~\ref{fig:RvW100}~ is similar to Fig.~\ref{fig:RvW500}, except that the benchmark point is $\lambda=100, \Omega_0=1/10$. As with the previous plot the difference between the numeric and analytic results regarding the radius of the Q-balls is completely negligible. It is in principle possible to modify our radius prediction to obtain better accuracy for even larger AdS curvature. Any miscalculation of the radius feeds into the profile prediction and, in fact, is the largest source of error for near thin-wall profiles. 
As seen in the right panel of Fig.~\ref{fig:RvW100} when we correct for the radius the functional form of the analytic thin-wall profile remains impressively accurate. The profile with smaller radius is not fit well by our analytic profile. This is because our calculation assumed the particle begins rolling from the maximum of the potential, but for these profiles the particle begins rolling some way downhill of the maximum.

 \begin{figure}
\centering
  \includegraphics[width=.8\linewidth]{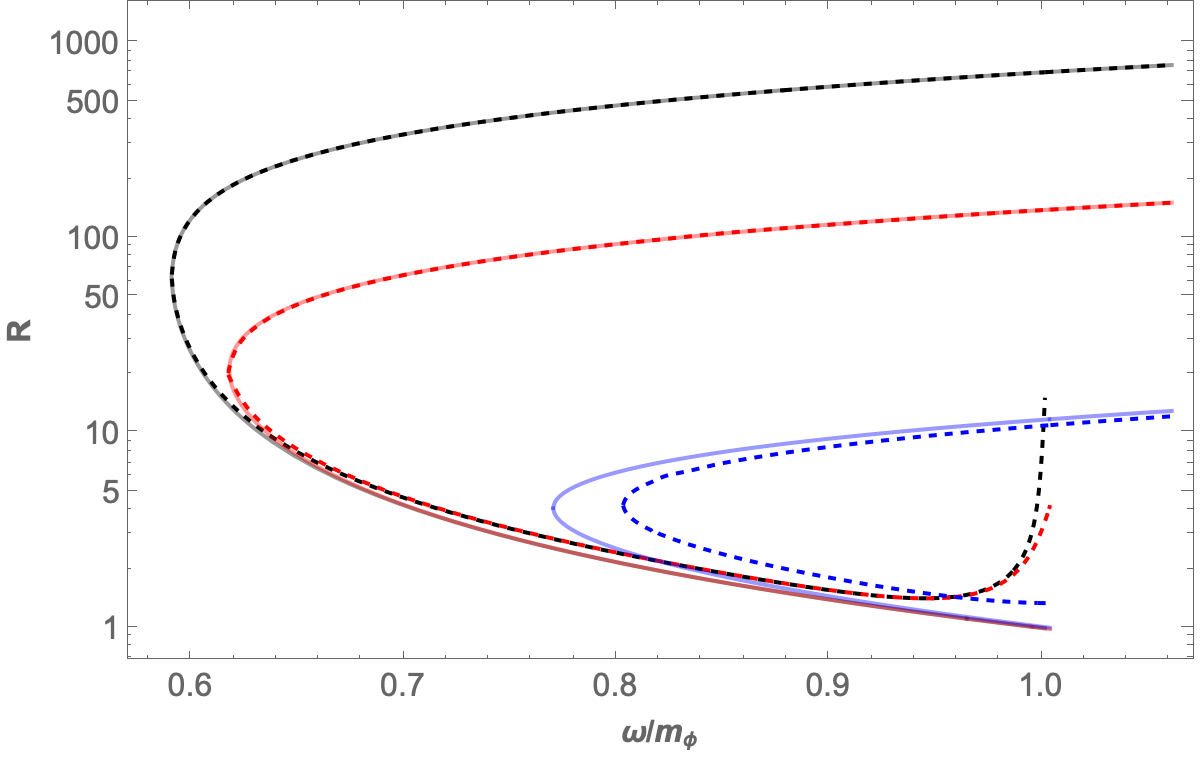}
\caption{Plot of radius vs $\omega/m_\phi$ for $\lambda$ = 500 (black), 100 (red), and 10 (blue). The theoretical predictions for the radius are shown in solid lines while the numerical data are show as dashed lines.}
\label{fig:ChangeLambda}
\end{figure}

As can be seen, there is overall an excellent match between the analytical predictions from the previous section and the numerical results near the thin-wall limit. Away from the thin-wall limit the analytic approximations do not provide a good fit to the profile. It is likely that this fit could be improved by making a more carful analysis of profiles that do not begin near the maximum of the potential. 

In Fig.~\ref{fig:ChangeLambda} we plot the predicted (solid) and numerical (dashed) radius for AdS Q-balls for $\lambda=$ 500 (black), 100 (read), and 10 (blue). A few general characteristics of the solution space are evident. For larger and larger $\lambda$ we approach the flat space limit, in which the upper branch of solutions is pushed to infinite radius. As the $\lambda$ is increased the value of $\omega$ at which the upper branch begins becomes larger, leading to a smaller radius at which $\omega=m_\phi$. We also see that the lower branch Q-balls near $\omega=m_\phi$ with increasing radius are seemingly pulled out of the solution space as $\lambda$ increases. 

These observations are quite similar to what has been demonstrated for gauged Q-balls in flat space~\cite{Lee:1988ag}. In that case, increasing the gauge coupling has a similar effect to increasing $\lambda$~\cite{Gulamov:2015fya}. This can be understood by the mapping of global Q-balls into the gauged Q-ball solution space~\cite{Heeck:2021zvk,Heeck:2021gam}, which suggests that even though flat-space gauged Q-balls differ significantly from global AdS Q-balls that a similar mapping from the flat space global solution to the global AdS solution space can also be defined.

We emphasize that from the figures, unlike in flat space~\cite{Coleman:1985ki}, there are AdS Q-balls for $\omega>m_\phi$. We have shown this for the three benchmarks in Fig.~\ref{fig:ChangeLambda}, but it seems to apply quite generally. As discussed in the previous section, in the full profile equation (\ref{eq:eom}), the term proportional to $\Omega$ is itself suppressed at large $\rho$, and so we can have solutions that are localized, even if $\omega>m_\phi$. This might appear to suggest that these AdS Q-balls have an arbitrarily large radius. The following sections, however, argue that this is not the case.

\section{Instabilities\label{s.instab}}
In this section we discuss two different instabilities related to global Q-balls. The first is familiar from flat-space solitons, but the second is new to the AdS case.

(a) There exist solutions to the field equation with $E/Q>m_\phi$ and $\frac{dE}{dQ}<m_\phi$, where $E$ is the energy, and $Q$ is the global charge, for both flat-space and AdS Q-balls. At these points in parameter space a Q-ball is unstable to dissociation into $Q$ individual, independent particles of mass $m_\phi$. The emission of just a few particles does not stabilize the soliton because $\frac{dE}{dQ}=\omega<m_\phi$, so shedding individual charged particles makes $E/Q$ even larger. Therefore, any soliton of this type completely falls apart. Thus, the lowest energy state with this charge is expected to be a gas of free particles without any soliton contribution. 

(b) In AdS, though not in flat space, we can have $\frac{dE}{dQ}>m_\phi$. In this case, removing a particle from the Q-ball decreases the energy of the system, as the energy required to remove the particle is more than compensated by the binding energy.
Such a Q-ball will therefore radiate particles until it reaches a state with  $\frac{dE}{dQ}=m_\phi$, with an additional
gas of free particles of mass $m_\phi$.
From the relation $\frac{dE}{dQ}=\omega$, we find that if for a global charge $Q$ the Q-ball solution has $\omega<m_\phi$ it remains a pure Q-ball solution. Conversely, if for a fixed charge $Q$ the soliton solution has $\omega>m_\phi$ it sheds particles leading to a field configuration with a Q-ball plus additional unbound particles.

These instability points are marked for the benchmark points in Figs.~\ref{fig:RvW500} and~\ref{fig:RvW100}. The red dot indicates the point at which $E=m_\phi Q$; to the right of this point, the lower branch becomes unstable to a state completely composed of free particles. We have also shown the line $\omega=m_\phi$. To the right of this line we have ${dE\over dQ}>m_\phi$ and the upper branch becomes unstable to a soliton of smaller charge surrounded by a gas of free particles. 

The implication is that in AdS generically, there is, first, a minimum
charge $Q_\text{min}$ for Q-balls below which $E=m_\phi Q$ and the soliton dissolves into particles. Second, a maximum
charge $Q_\text{max}$ for Q-balls where $\omega =m_\phi$, and beyond which a gas of particles forms around the Q-ball. Third, the two branches of solutions as function of $\omega$ shows that there is a charge $Q$ where the frequency $\omega$ is minimized to some value $\omega_{min}$.

\begin{figure}
  \includegraphics[width=.8\linewidth]{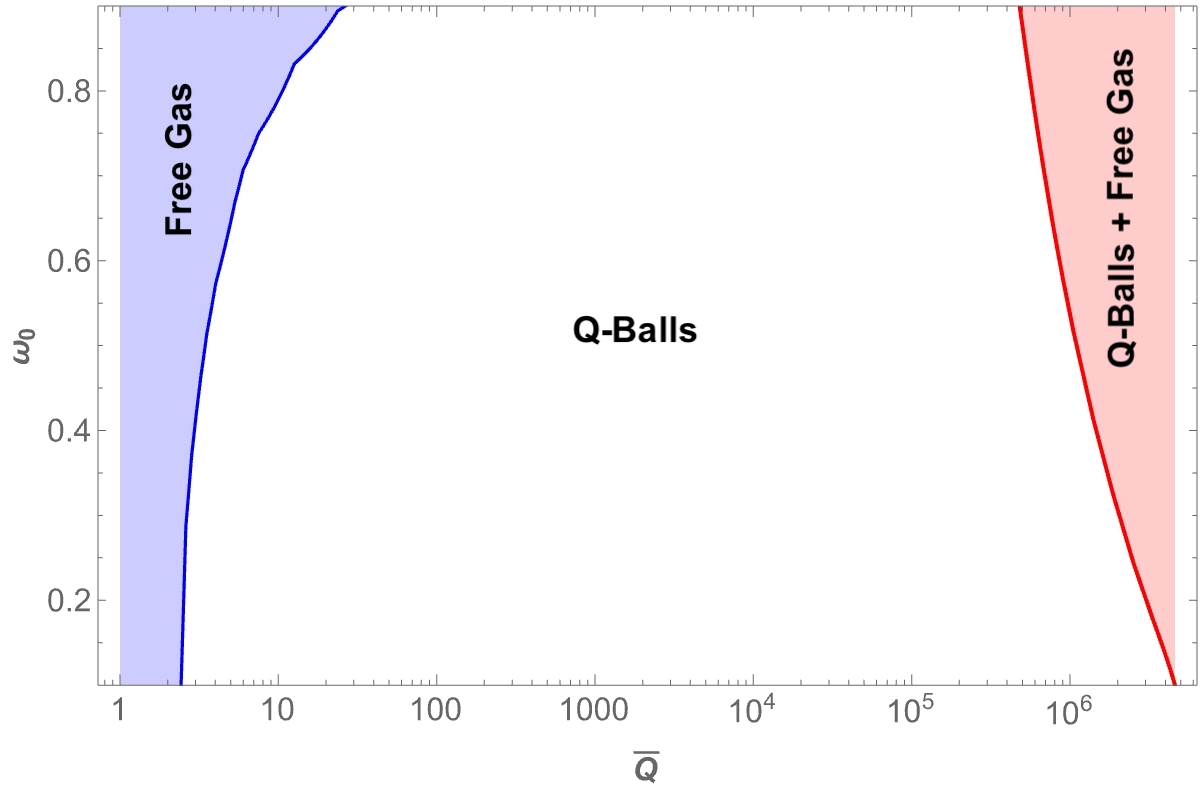}
\caption{
Phase diagram of the theory as a function of $\omega_0$ and the total charge $\bar{Q} \equiv Q/4\pi\phi_{0}^{2}$   for $\lambda = 100$. The phase on the right (red) appears for Q-balls in the AdS background, but not in flat space.}
\label{fig:phases}
\end{figure}

In Fig.~\ref{fig:phases} we illustrate the phase diagram of AdS Q-balls as a function of the total charge $Q$ and $\omega_0$ for the benchmark value of $\lambda=100$. The phase of a free gas of unbound particles is largely the same as for flat space Q-balls. However, the phase on the right of the diagram cannot occur in flat space.

\section{Comments on large Radius AdS Q-balls\label{s.LR}}
It is interesting to note that in AdS space, the thin-wall Q-balls must have a maximal radius. This is 
easily seen by using the language of a particle rolling in a potential.
In the thin-wall approximation, the field stays at the second maximum till it transitions. However, for extremely large $R$, the energy of the second maximum drops below the energy at $f=0$, and the transition can no longer occur. There is therefore a maximum possible radius.

There are two possible avenues to having stable AdS Q-balls with large radii. First, one could 
consider non-thin-wall solutions where the Q-ball profile rolls slowly between the two maxima. 
These solutions cannot be treated analytically using our methods, and so the formula (\ref{eq:RvWreln}) does not apply. However, the general argument of the previous paragraph still applies; as $\rho$ increases, the entire potential drops, and eventually the field does not have the energy to return to the final field value at  $f=0$.

The other possibility is to consider an entirely different class of potential. The effective potential for the particle is of the form
\beq
V(f,\rho)={\Omega^2f^2\over \left(1+{\rho^2\over \lambda^2}\right)^2}-\frac{U(f)}{1+{\rho^2\over \lambda^2}}~.
\eeq
For large $\rho$, the first term is additionally suppressed, and in our case, we recover similar dynamics to the original sextic potential, where the second maximum is below the maximum at $f=0$. 
However, we can consider a potential where the first term is always dominant at large $f$; this can happen if $U(f)$ grows slower than $f^2$ at large $f$. An example is the exponential potential considered in~\cite{Hartmann:2012wa}, where the potential goes to a constant at large $f$. In this model, the authors of~\cite{Hartmann:2012wa} indeed found Q-balls of arbitrarily large radius and charge. We therefore find that  the existence of large Q-balls in AdS is possible, but requires a specific form of potential. We leave this issue for future consideration.

\section{Implications in the dual theory\label{s.CFT}}
From the AdS/CFT correspondence, we expect our results to have an interpretation in a dual three-dimensional field theory.
While the precise nature of this field theory is not determined, we note that
the scalar field in the AdS bulk must map to a scalar field operator in the dual boundary theory. This operator is typically a composite of the fundamental fields of the theory. We leave the nature of the fundamental theory unspecified, but assume that it has a low energy limit or sector which is dominated by the interactions of this scalar operator. One intriguing possibility is that the AdS Q-balls are dual to scar states~\cite{Serbyn:2020wys,Cotler:2022syr}. These states have been tied to nontopological AdS soliton states without horizons~\cite{Milekhin:2023was}, such as boson stars and oscillons, though the connections to Q-balls has not been established. In any case, the bulk potential determines certain properties of the scalar operator, and in particular, the dimension of this operator is~\cite{Aharony:1999ti}
\bea
\Delta ={d\over 2}+\sqrt{{d^2\over 4}+\ell^2m_\phi^2}~,
\label{dimeq}
\eea
where $d$ is the dimension of the space and $\ell$ is the AdS radius.

The charge of the $U(1)$ scalar field in AdS maps to a $U(1)$ charge on the boundary. More precisely, by weakly gauging the $U(1)$ global symmetry of bulk theory (so that the gauge field does not affect the structure of the Q-ball) this symmetry would be dual to a global $U(1)$ symmetry on the boundary.
By construction, the scalar field operator on the boundary carries a charge under this global $U(1)$,  
and states created by the action of this operator also carry this global charge.

We can understand the dynamics of this sector, and in particular the structure of the ground state of the theory as a function of the 
charge $Q$, using the bulk Q-ball solution.

{\it Small charge}: From the bulk description, we see that for very small charges, the solution is a gas of free particles. The energy increases proportionally to $Q$ as $E=m_\phi Q$ (here $m_\phi$ is related to the operator dimension by \ref{dimeq}).

{\it Intermediate charge}:  As we increase the charge, there is a critical value (corresponding to $Q_\text{min}$) at which the ground state becomes a condensate carrying charge $Q$. The Q-ball is localized in the interior of the AdS space, and has a nontrivial dependence on the $r$ coordinate. This indicates that the three dimensional field theory condensate has a nontrivial dependence on scale.

At the transition point, the condensate has an energy $E=m_\phi Q$. The transition therefore does not lead to a discontinuous jump in the energy. However, the condensate has an energy dependence of the form
${dE\over dQ}=\omega$, in contrast to the free gas which has the  relation ${dE\over dQ}=m_\phi$.
Since the transition occurs at a value $\omega$ which is not equal to $m_\phi$ (as seen in Figs.~\ref{fig:RvW500} and~\ref{fig:RvW100}), we see that there is a discontinuity in   ${dE\over dQ}$. This is therefore a second order phase transition. As the charge increases further, the condensate becomes increasingly bound till
the binding energy per particle reaches a maximum value of $m_\phi-\omega_\text{min}$. 

{\it Large charge}: As we continue increasing the charge, the binding energy begins to decrease, and
  vanishes at $Q_\text{max}$. Beyond this point, additional charge is not absorbed into the condensate.
However, the condensate does not evaporate; instead the condensate is surrounded by a gas of free particles. More precisely, the condensate carries a charge $Q_\text{max}$, and the remaining charge  
$Q-Q_\text{max}$ is in free particles.

As mentioned above, the condensate has an energy dependence of the form
${dE\over dQ}=\omega$. Since the transition occurs exactly at $\omega=m_\phi$, the energy dependence is ${dE\over dQ}=m_\phi$. This is equal to the change in energy when we add a free particle. This implies that ${dE\over dQ}$ is continuous.
However, for the gas around the Q-ball, ${d^2E\over dQ^2}=0$, while for the Q-ball 
${d^2E\over dQ^2}={d\omega\over dQ}$.
Hence ${d^2E\over dQ^2}$ is discontinuous, which implies that this is a \emph{third-order} phase transition.

\section{Q-balls in n-dimensional Anti-de Sitter\label{s.nDimAdsQ}}
\begin{figure}
\centering
\begin{subfigure}{.5\textwidth}
  \centering
  \includegraphics[width=.8\linewidth]{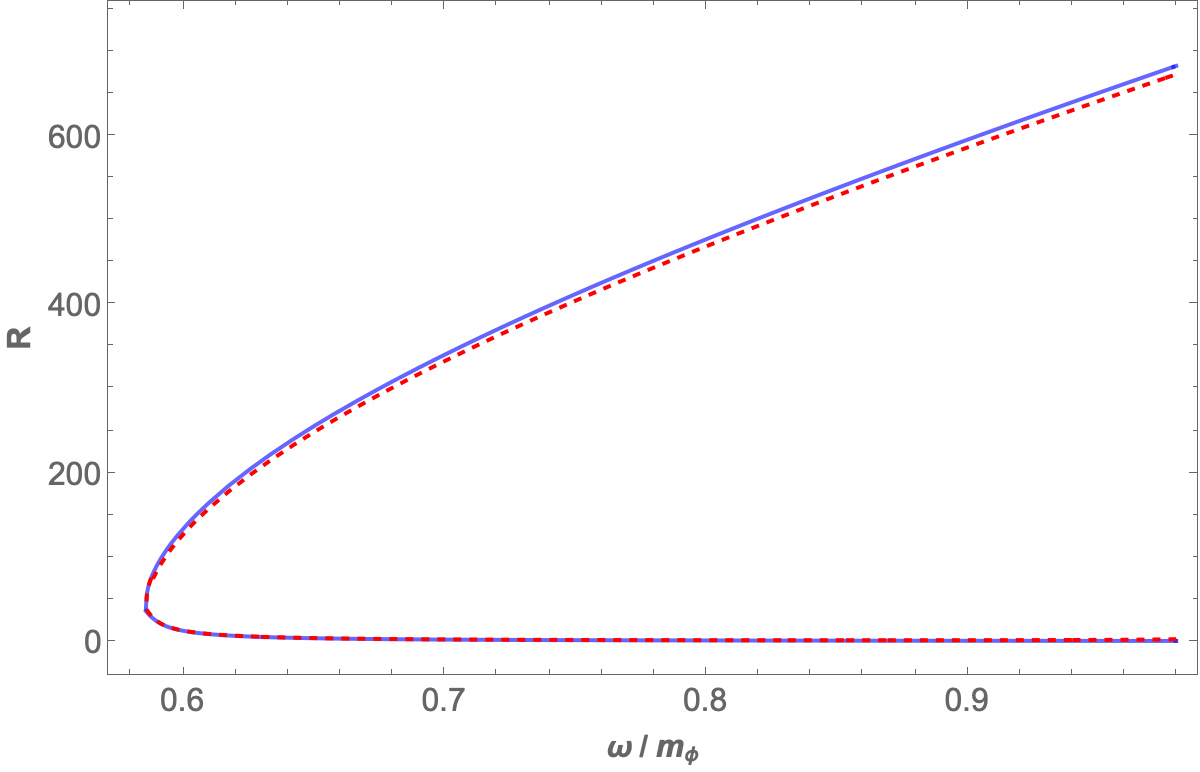}
  \label{fig:sub1_3}
\end{subfigure}%
\begin{subfigure}{.5\textwidth}
  \centering
  \includegraphics[width=.8\linewidth]{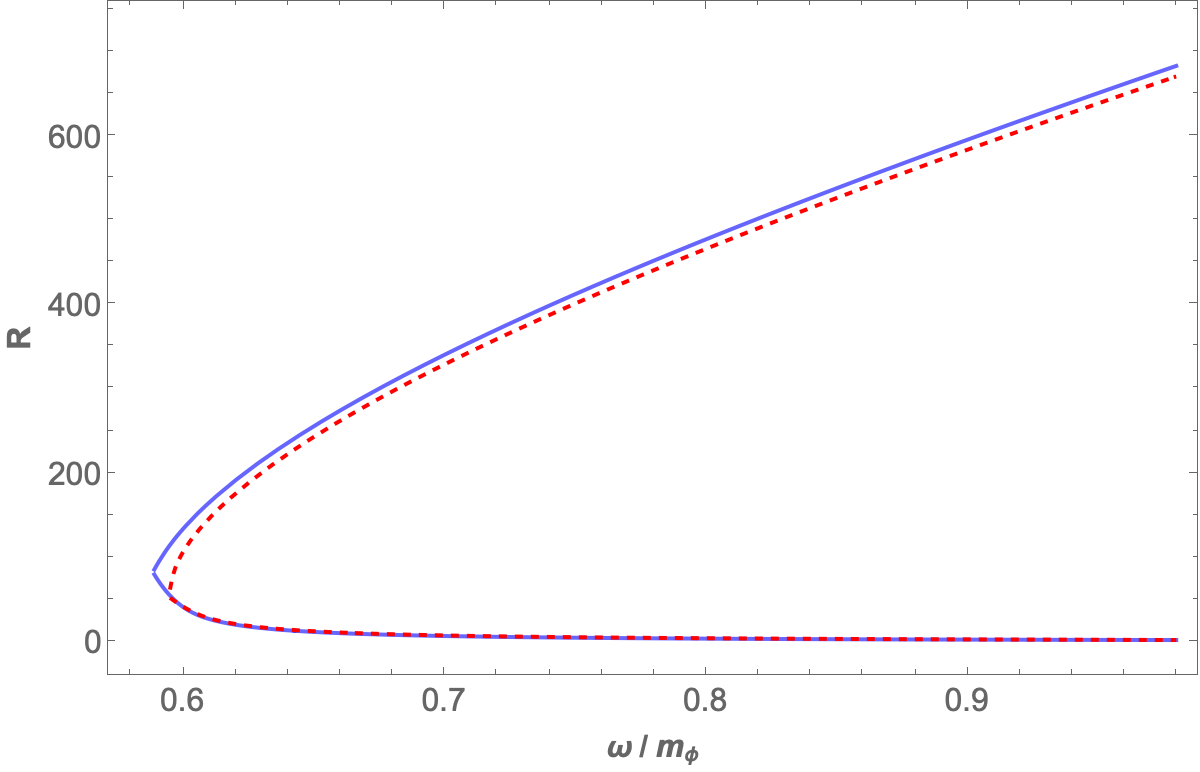}
  \label{fig:sub2_3}
\end{subfigure}
\caption{
Numerical calculations (red dashed) compared to analytic predictions (blue) for $\lambda_n=500,\, \Omega_0=1/\sqrt{2}$. Left panel: Q-ball radius $R$ as a function of $\omega$ for $n=3$. Right panel: Q-ball radius $R$ as a function of $\omega$ for $n=5$.}
\label{fig:n3n5Rvw}
\end{figure}

The above analysis has assumed a 3+1 AdS space which maps to a 2+1 CFT. In this section we outline how our results generalize to $(n-1)+1$ AdS spaces related to $(n-2)+1$ CFTs. 
The AdS geometry in $n$-dimensions is parameterized as
\beq
ds^2=a(r)dt^2-b(r)dr^2-r^2d\Omega_{n-2}^2~,
\eeq
with
\beq
a(r)=1+\frac{r^2}{\ell_n^2}~, \ \ b(r)=\frac{1}{1+\frac{r^2}{\ell_n^2}}~,\label{e.nDimAdSgrav}
\eeq
where $\ell_n = \sqrt{-\frac{(n-2)(n-1)}{\Lambda}}$ and $d\Omega_{n-2}^{2}$ is the measure on the $(n-2)$-sphere. Assuming a spherically symmetric solution as before, the action becomes
\beq
\mathcal{S} = S_{n-2}\phi_{0}^{2}\int dt\ dr\ r^{(n-2)}\left[ \frac{1}{2}\frac{\omega^{2}f^{2}}{1+r^{2}/\ell_{n}^{2}} - \frac{1}{2}f'^{2} (1+r^{2}/\ell_{n}^{2})-\overline{U}(\phi_{0}^{2}f^{2}/2)\right],
\eeq
where
\beq
\overline{U} = \frac{1}{2}(m_{\phi}^{2}-\omega_{0}^{2})f^{2}(1-f^{2})^{2}+\frac{\omega_{0}^{2}}{2}f^{2}.
\eeq
We can then consider $f$ to be a particle rolling in a time dependent potential, as before, with Lagrangian
\beq
L = r^{(n-2)} \left[ \frac{1}{2}\frac{\omega^{2}f^{2}}{1+r^{2}/\ell_{n}^{2}} - \frac{1}{2}f'^{2} (1+r^{2}/\ell_{n}^{2})-\frac{1}{2}(m_{\phi}^{2}-\omega_{0}^{2})f^{2}(1-f^{2})^{2}-\frac{\omega_{0}^{2}}{2}f^{2}\right].
\eeq
Transforming to dimensionless parameters $\rho$, $\Omega$, $\Omega_{0}$, and $\lambda_{n} = \ell_{n}\sqrt{m_{\phi}^{2}-\omega_{0}^{2}}$, the Lagrangian can be written as
\beq
L = \frac{\rho^{(n-2)}}{(m_{\phi}^{2}-\omega_{0}^{2})^{n-3}}\left[ \frac{1}{2}\frac{\Omega^{2}f^{2}}{1+\rho^{2}/\lambda_{n}^{2}}-\frac{1}{2}(f')^{2}(1+\rho^{2}/\lambda_{n}^{2})-\frac{1}{2}f^{2}(1-f^{2})^{2}-\frac{\Omega_{0}^{2}}{2}f^{2} \right].
\eeq
The equation of motion for $f(\rho)$ is
\begin{align}
f''+\frac{n-2}{\rho}&\frac{1+(\frac{n}{n-2})(\rho^{2}/\lambda_{n}^{2})}{1+\rho^{2}/\lambda^{2}_{n}}f' \nonumber \\
&=\frac{f}{1+\rho^{2}/\lambda_{n}^{2}}\left[1+\Omega_{0}^{2}-\frac{\Omega^{2}}{1+\rho^{2}/\lambda_{n}^{2}}-4f^{2}+3f^{4}\right] \equiv -\frac{\partial V}{\partial f},\label{eq:nDimfEqn}
\end{align}
where $V(f,\rho)$ is the effective potential, given explicitly by
\beq
V(f,\rho)=\frac{f^2}{2\left(1+\rho^2/\lambda_{n}^2 \right)}\left[\frac{\Omega^2}{1+\rho^2/\lambda_{n}^2}-\Omega_0^2-(1-f^2)^2 \right]
\eeq
after imposing $V(0,\infty)=0$. The effective potential is unchanged from the $n=4$ case aside from the substitution $\lambda \rightarrow \lambda_{n}$. Defining the energy-like quantity $W$ as in Eq. (\ref{eq:edef}), we find $W$ is approximately conserved for the thin-wall solutions, and the approximate thin-wall profile for the $n$-dimensional case is then given by
\beq
f_{T,n}(\rho) = \frac{f_{+,n}(\rho)}{\sqrt{1+2e^{2(\rho-R)/\sqrt{1+R^{2}/\lambda_{n}^{2}}}}},\label{eq:NDimTrans}
\eeq
with
\beq
f_{+,n}(\rho) \equiv \frac23 + \frac13\sqrt{1+\frac{3\Omega^2}{1+\rho^2/\lambda_n^2}-3\Omega_0^2}.
\eeq
Using the equations of motion for $f$, Eq. (\ref{eq:nDimfEqn}), we find
\bea
{dW\over d\rho}=-
\frac{(n-2)}{\rho}\frac{1+(\frac{n}{n-2})\rho^2/\lambda_{n}^2}{1+\rho^2/\lambda_{n}^2}f^{\prime2}
+\frac{\rho f^2}{\lambda_{n}^2\left(1+\rho^2/\lambda_{n}^2 \right)^2}\left[\Omega_0^2-\frac{2\Omega^2}{1+\rho^2/\lambda_{n}^2 }+\left(1-f^2\right)^2 \right].
\eea
By integrating from $\rho=R-z_0$ to $\rho=\infty$ we find
\begin{align}
W_{\infty}-W_{R-z_0}=&-
(n-2)\int_{R-z_0}^{\infty}\frac{d\rho}{\rho}\frac{1+(\frac{n}{n-2})\rho^2/\lambda_n^2}{1+\rho^2/\lambda_n^2}f^{\prime2}
\nonumber
\\
&+\int_{R-z_0}^{\infty}d\rho\frac{\rho f^2}{\lambda_n^2\left(1+\rho^2/\lambda_n^2 \right)^2}\left[\Omega_0^2-\frac{2\Omega^2}{1+\rho^2/\lambda_n^2 }+\left(1-f^2\right)^2 \right].
\label{eq:integralNDim}
\end{align}
Once again, we compute the leading order effect of the friction by evaluating these integrals using the transition function defined in Eq. (\ref{eq:NDimTrans}), this results in the relation
\beq
\Omega_{0}^{2} - \frac{\Omega^{2}}{1+\frac{R^{2}}{\lambda_{n}^{2}}} = \frac{\lambda_{n}^{2}R^{2}(\Omega^{2}\ln(4)-2n+3)-(n-2)\lambda_{n}^{4}-(n-1)R^{4}}{2R\lambda_{n}(\lambda_{n}^{2}+R^{2})^{3/2}}~.
\label{eq:ndimRadPredic}
\eeq
This relation is derived in more detail in Appendix \ref{app:Rderiv}. Note that this relation exactly matches the previous implicit equation for $R$, Eq. (\ref{eq:RvWreln}), for $n=4$. Fig. \ref{fig:n3n5Rvw} shows the numerical and predicted Q-ball radius as a function of $\omega$ for $n=3$ and $n=5$, with $\lambda_n=500$ and $\Omega_{0} = 1/\sqrt{2}$ for both cases. These results show that AdS Q-balls can be constructed in various dimensions by these methods. Consequently, the qualitative analysis of their dual CFTs follows, including similar looking phase diagrams to what was shown in the previous section. 

\section{Discussion and Conclusion\label{s.Conc}}
In this article we have studied Q-balls in a fixed AdS background, with particular focus on thin-wall Q-balls. The scalar field is subject to a sextic potential (which also supports Q-balls in a flat background), but the general features are expected to apply most, if not all, potentials that support thin-wall solitons. We have shown that the AdS background significantly modifies the physics of these Q-balls, even if gravity is not dynamical. In particular, Q-balls have a maximal radius in AdS, unless the potential is of a very special type. 

We have also found new phases of the scalar field that composes the Q-balls which do not exist in flat space. Specifically, for large charges, the lowest energy field configuration is a Q-ball surrounded by a gas of massive particles, while in flat space, the ground state for larger charge is just a Q-ball with larger mass and radius.

The transitions between the states are also of a novel type. At low charges, as the charge is increased, there is a zero-temperature second order phase transition to a Q-ball state, similar to flat space systems. Curiously, at high charges, there is a {\it third}-order phase transition to the mixed Q-ball scalar gas state mentioned above. This phase transition has no flat space analogue. These results imply that the dual boundary theory must also have similar transitions. It would be quite interesting to explore whether scar states, similar to those related to other non-topological solitons in AdS, can be related to AdS Q-balls. If so, then one should be able to understand this third order phase transition in the context of scar states.

In this work we have ignored backreaction on the geometry. This implies that the gravitational interactions of order ${1\over M_\text{pl}}$ have been dropped, while keeping the curvature scale of the AdS constant. In terms of the original AdS/CFT correspondence, this amounts to taking the $N$ of the dual gauge group to infinity while keeping the t'Hooft coupling $g^2N$ fixed.

Including dynamical gravity on the AdS side would be an important next step to see if these novel phase transitions survive the inclusion of gravity. It would also be interesting to connect these solutions to the known boson star solutions. Finally, there is cosmologically motivation to extend these solutions to a full characterization of Q-balls in de Sitter space~\cite{Palti:2004is}. We leave these and other questions for future work.

\section*{Acknowledgements}
We thank Julian Heeck for helpful comments on various aspects of this work. The work of  A.R. is supported by the National Science Foundation Grant No.~PHY-1915005. C.B.V. is supported in part by the National Science Foundation under Grant No. PHY-2210067.

\appendix

\section{Derivation of Q-Ball Radius Formula}\label{app:Rderiv}
In order to derive Eq. (\ref{eq:ndimRadPredic}), we must evaluate
\begin{align}\label{eq:rad_relation_app}
W_{\infty}-W_{R-z_0}=&-
(n-2)\int_{R-z_0}^{\infty}\frac{d\rho}{\rho}\frac{1+(\frac{n}{n-2})\rho^2/\lambda_n^2}{1+\rho^2/\lambda_n^2}f^{\prime2}
\nonumber
\\
&+\int_{R-z_0}^{\infty}d\rho\frac{\rho f^2}{\lambda_n^2\left(1+\rho^2/\lambda_n^2 \right)^2}\left[\Omega_0^2-\frac{2\Omega^2}{1+\rho^2/\lambda_n^2 }+\left(1-f^2\right)^2 \right].
\end{align}
We assume $R \gg \lambda_{n}$. The first integral on the right hand side can be approximated by noting that the dominant contribution occurs about the transition between minima at $\rho = R$. Define the function $f_{t,n}(\rho)$ as
\begin{equation}
	f_{t,n}(\rho) = \frac{1}{\sqrt{1+2e^{2(\rho-R)/\sqrt{1+R^{2}/\lambda_{n}^{2}}}}},
\end{equation}
so that $f(\rho) = f_{+,n}(\rho)f_{t,n}(\rho)$. Around $\rho = R$, we have $f'(\rho) \approx f_{+,n}(\rho)f'_{t,n}(\rho)$. Since the integrand is exponentially suppressed below $\rho = R$, we can integrate from $\rho = 0$ to $\infty$ rather than from $R-z_{0}$ to $\infty$. This integral can be performed analytically, and after dropping another exponentially suppressed term in the result it yields:
\begin{equation}\label{eq:int1_app_A}
	(n-2)\int_{R-z_0}^{\infty}\frac{d\rho}{\rho}\frac{1+(\frac{n}{n-2})\rho^2/\lambda_n^2}{1+\rho^2/\lambda_n^2}f^{\prime2} \approx-\frac{\lambda_{n}\big(\lambda_{n}^{2}(n-2)+nR^{2}\big)\left( 2+\sqrt{1+\frac{3\lambda_{n}^{2}\Omega^{2}}{\lambda_{n}^{2}+R^{2}}-3\Omega_{0}^{2}} \right)}{12R(\lambda_{n}^{2}+R^{2})^{3/2}}.
\end{equation}

The second integral in Eq. (\ref{eq:rad_relation_app}) can be evaluated by expanding the integrand about $\rho=R$ to first order in $\rho-R$ (except the exponential piece in $f_{t,n}(\rho)$ since its series expansion does not converge; however, we will be able to do these integrals analytically still while including this exponential piece). The resulting expression will be very large and have a dependence on $z$, but this dependence will match exactly with the $z$ dependence on the left hand side of Eq. (\ref{eq:rad_relation_app}) and cancel out.

On the left hand side we need to expand $W(\rho)$, given by Eq. (\ref{eq:edef}), about $\rho = R-z$; the $W_{\infty}$ piece is simply zero since the profile and potential vanish at infinity. It will be convenient to instead write $\rho = R(1-a)$ where $a = z/R$ and expand about $a = 0$. Once again, we do not expand the exponential part of $f_{t,n}(\rho)$ since this does not converge, but we will find this just leads to extra exponentially suppressed terms which can be dropped. After doing this substitution and expansion, we can send $a \rightarrow z/R$ and then expand the resulting expression about $z=0$. Once again, this results in a very large expression with some $z$ dependence.

Noting $W_{\infty} = 0$, we can rewrite Eq. (\ref{eq:rad_relation_app}) as
\begin{align}\label{eq:rad_rel_rewrite}
	&W_{R-z_0}-
(n-2)\int_{R-z_0}^{\infty}\frac{d\rho}{\rho}\frac{1+(\frac{n}{n-2})\rho^2/\lambda_n^2}{1+\rho^2/\lambda_n^2}f^{\prime2}\nonumber\\
&+\int_{R-z_0}^{\infty}d\rho\frac{\rho f^2}{\lambda_n^2\left(1+\rho^2/\lambda_n^2 \right)^2}\left[\Omega_0^2-\frac{2\Omega^2}{1+\rho^2/\lambda_n^2 }+\left(1-f^2\right)^2 \right]=0.
\end{align}
Plugging in our expressions for each term and expanding in powers of $z$, we will find that the $z$ dependence exactly cancels out to second order. To find an implicit expression for the Q-ball radius to leading order in $1/R$, we first rewrite $\lambda_{n}$ in terms of a new free parameter $\sigma$ via $\lambda_{n} = \sigma R$. We then search for an equation of the form
\begin{equation}\label{eq:ndim_R_implicit}
	\Omega_{0}^{2} - \frac{\Omega^{2}}{1+\sigma^{-2}} = f_{0}(n,\sigma) + \frac{1}{R}f_{1}(n,\sigma),
\end{equation}
which mirrors the form of the radius relation in~\cite{Heeck:2020bau} with $1/R$ corrections. Enforcing this relation on Eq. (\ref{eq:rad_rel_rewrite}) we find
\begin{equation}
	f_{0}(n,\sigma) = 0, \quad f_{1}(n,\sigma) = \frac{1+2\sigma^{4} - n(1+\sigma^{2})^{2} + \sigma^{2}(3+\Omega^{2}\ln(4))}{2\sigma (1+\sigma^{2})^{3/2}}.
\end{equation}
Plugging these results into Eq. (\ref{eq:ndim_R_implicit}) and sending $\sigma \rightarrow \lambda_{n}/R$ reproduce Eq. (\ref{eq:ndimRadPredic}).

\bibliographystyle{utcaps_mod}
\bibliography{BIB}

\end{document}